\documentclass[a4paper,11pt]{article}
\usepackage{pos}

\newcommand{\raa}{\ensuremath{R_{\mathrm{AA}}}}
\newcommand{\pt}{\ensuremath{p_{\mathrm{T}}}}
\newcommand{\cmet}{\ensuremath{\sqrt{s} = 5.02 } TeV}

\newcommand{\cmetthirteen}{\ensuremath{\sqrt{s} = 13 } TeV}
\newcommand{\cmetnn}{\ensuremath{\sqrt{s_{\mathrm{NN}}} = 5.02 } TeV}
\newcommand{\dzero}{\ensuremath{\mathrm{D}^0}}
\newcommand{\dplus}{\ensuremath{\mathrm{D}^+}}
\newcommand{\dstar}{\ensuremath{\mathrm{D}^{*+}}}
\newcommand{\ds}{\ensuremath{\mathrm{D}_{\mathrm{s}}^{+}}}
\newcommand{\lc}{\ensuremath{\mathrm{\Lambda}_{\mathrm{c}}^{+}}}
\newcommand{\epm}{\ensuremath{\mathrm{e}^{\pm}}}
\newcommand{\gevc}{\ensuremath{\ \mathrm{GeV}/c}}

\title{Open heavy-flavour production from small to large collision systems with ALICE at the LHC}
\ShortTitle{Open heavy-flavour production with ALICE at the LHC}

\author*[\dag]{Fabio Catalano}

\affiliation{Politecnico di Torino,\\
             Corso Duca degli Abruzzi 24, Torino, 10129, Italy}
\affiliation{INFN sez. Torino,\\
                Via Pietro Giuria 1, Torino, 10125, Italy}

\notes{\note{on behalf of the ALICE Collaboration}}

\emailAdd{fabio.catalano@cern.ch}

\abstract{Heavy quarks are effective probes of the hot and dense nuclear matter, the quark--gluon plasma, produced in ultra-relativistic heavy-ion collisions. Due to the short time scale characterising their production, heavy quarks experience the whole evolution of the system. In particular, measurements of heavy-flavour hadron production in Pb--Pb collisions at LHC energies give insight into the mechanisms of heavy-quark transport in the deconfined matter. 
In small hadronic systems, pp and p--Pb collisions, heavy-flavour measurements provide the baseline for observations of hot-medium effects in heavy-ion collisions, as well as tests of perturbative quantum chromodynamic calculations and measurements of cold-nuclear-matter effects. In this contribution, recent ALICE results on open heavy-flavour hadron production in pp, p--Pb and Pb--Pb collisions at various energies are discussed. New measurements are presented both for fully-reconstructed charmed hadrons and for single electrons from heavy-flavour hadron decays, also investigating the beauty-quark production via the measurement of D mesons and electrons from beauty-hadron decays.
}

\FullConference{%
  HardProbes2020\\
  1-6 June 2020\\
  Austin, Texas}

\begin{document}
\maketitle

\section{Introduction}
Due to their large masses, charm and beauty quarks (heavy quarks) are mainly produced in hard-scattering processes between partons of the colliding nucleons. Therefore, their production can be described by perturbative quantum chromodynamic (pQCD) calculations down to zero transverse momentum (\pt{}). In ultra-relativistic heavy-ion collisions extreme temperatures are reached and lattice QCD calculations predict a phase transition of nuclear matter to a colour-deconfined medium, called quark--gluon plasma (QGP). 
Heavy quarks are produced in the initial stages of the nucleus--nucleus (Pb--Pb) collision, before the QGP formation, and experience the full evolution of the system while propagating through the medium and strongly interacting with the QGP constituents. 
In Pb--Pb collisions the measurement of hadrons containing heavy quarks provides crucial information on the in-medium parton energy loss. The comparison of heavy-flavour and light-flavour hadrons gives insight into the colour-charge and quark-mass dependence of energy loss. Moreover, the hadronisation mechanism of heavy quarks in the medium can be investigated comparing the production of different heavy-flavour hadron species, e.g., charmed baryons and mesons, or hadrons with and without strange-quark content \cite{Kuznetsova:2006bh}. Measurements of open heavy-flavour hadron production in proton-proton collisions are relevant tests of pQCD model calculations.

In ALICE, charmed hadrons are reconstructed at midrapidity (\(|y| <  0.8 \)) via the hadronic decay channels: \(\dzero{} \rightarrow \mathrm{K}^- \pi^+\), \(\dplus{} \rightarrow \mathrm{K}^- \pi^+ \pi^+\), \(\dstar{} \rightarrow \dzero{} \pi^+ \rightarrow \mathrm{K}^- \pi^+ \pi^+\), \(\ds{} \rightarrow \phi \pi^+ \rightarrow \mathrm{K}^- \mathrm{K}^+ \pi^+\), \(\lc{} \rightarrow \mathrm{p} \mathrm{K}^0_{\mathrm{s}}\) and their charge conjugates. Particle candidates are built from pairs or triplets of tracks with the proper charge combination. Kinematic and geometrical selections on the displaced decay-vertex topology, together with particle identification, are applied to reduce the combinatorial background. The charmed-hadron raw yields are obtained from an invariant-mass analysis and the reconstruction efficiencies are estimated using Monte Carlo simulations \cite{Acharya:2019mgn}. 
In addition, heavy-flavour hadrons are studied through the measurement of electrons produced in their semi-leptonic decays. Electrons are identified at midrapidity using the information provided by ALICE central-barrel detectors \cite{Aamodt:2008zz}. The hadron contamination and electrons from non-heavy-flavour sources, mainly photon conversions and Dalitz decays of light neutral mesons, are subtracted from the measured inclusive yield, which is then corrected for the acceptance and selection efficiency \cite{Acharya:2019mom}. Exclusive measurements of prompt charmed hadrons, originated from the hadronisation of charm quarks produced in the initial collision, and non-prompt ones, which are produced from beauty-hadron decays, are possible thanks to the longer proper mean-life of hadrons containing beauty quarks. This allows the ALICE experiment to assess beauty-quark production through the measurement of non-prompt D mesons and electrons from beauty-hadron decays.

\section{Open heavy-flavour production in pp collisions}
\begin{figure}[hbt]
	\centerline{
		\includegraphics[height=5.8cm]{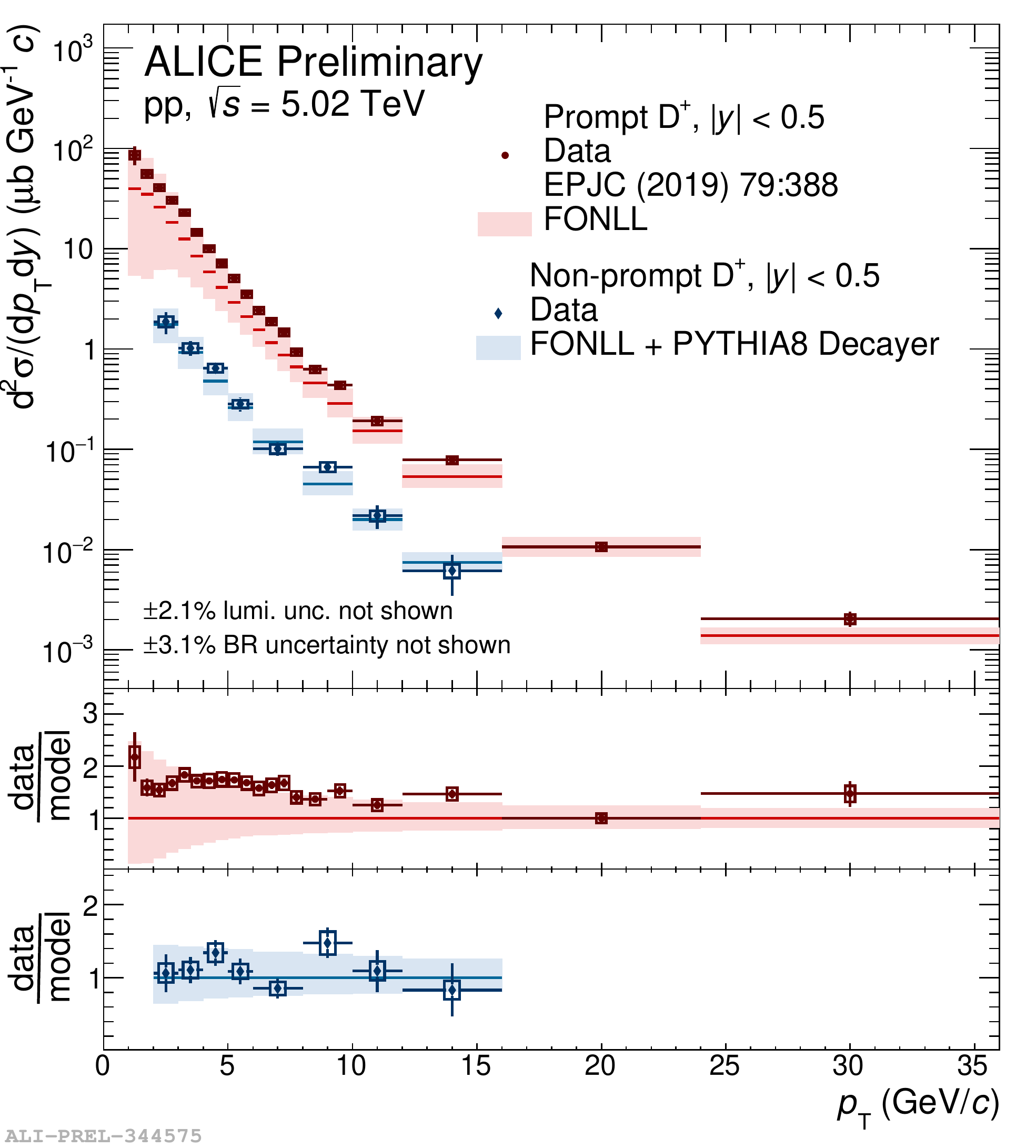}
    \includegraphics[height=5.8cm]{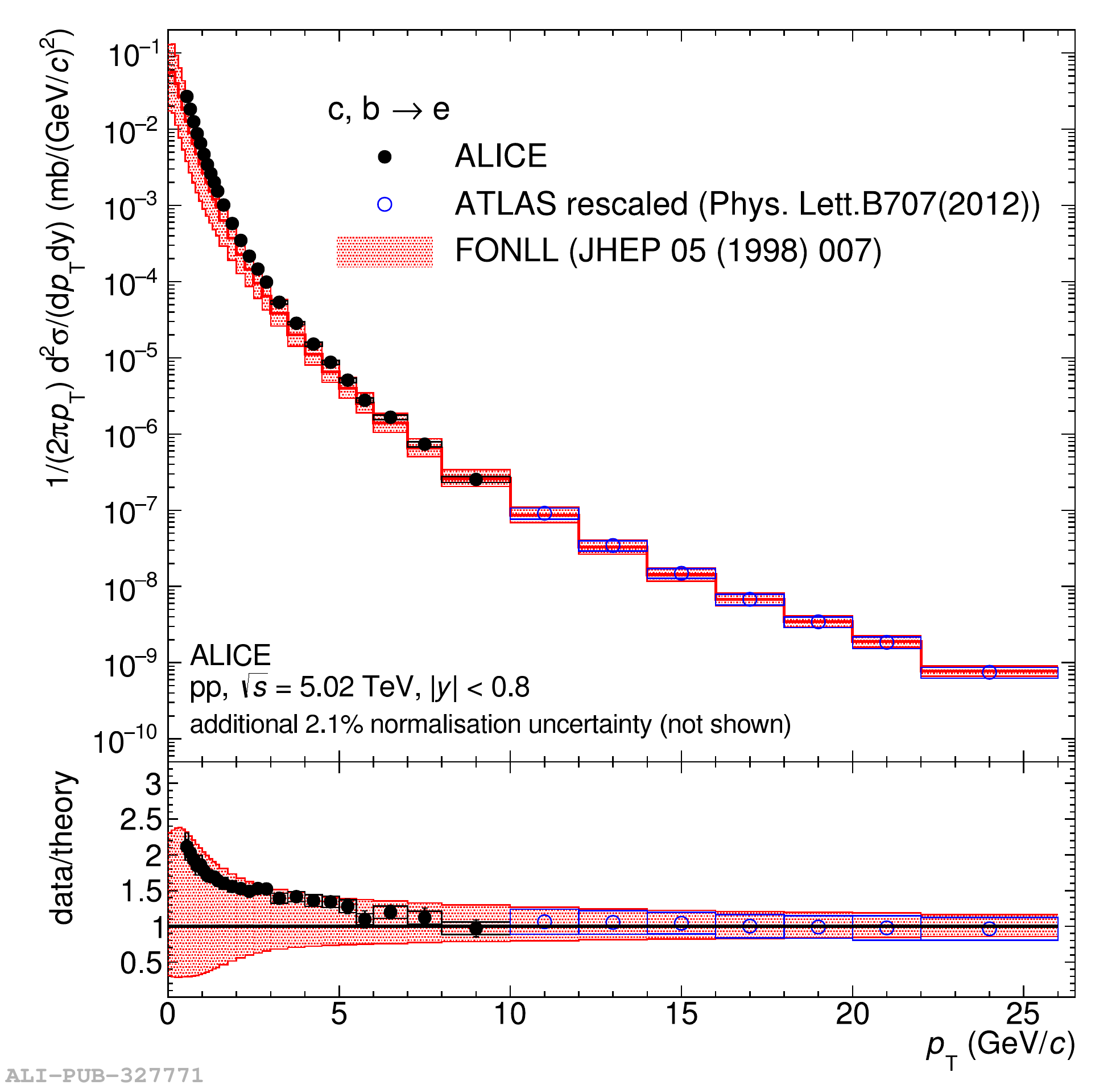}
	}
	\caption{Left: prompt and non-prompt \dplus{}-meson cross sections in pp collisions at \cmet{} compared to FONLL predictions. Right: \epm{} from heavy-flavour hadron decays compared to FONLL predictions. \label{pp}}
\end{figure}

The production cross section of prompt \cite{Acharya:2019mgn} and non-prompt D mesons and of electrons from semi-leptonic heavy-flavour hadron decays \cite{Acharya:2019mom} is measured at midrapidity in pp collisions at \cmet{}. In the left panel of Fig.~\ref{pp}, prompt and new measurements of non-prompt \dplus{} mesons are compared, respectively, to FONLL \cite{Cacciari:1998it}  predictions and to FONLL with the \(\mathrm{B} \rightarrow \mathrm{D} + \mathrm{X}\) decay kinematics described by the PYTHIA8 package \cite{Sjostrand:2014zea}. In the right panel, the cross section of heavy-flavour decay electrons is compared to theoretical calculations. 
The measurements are well described by pQCD calculations. The prompt \dplus{}-meson measurement lies on the upper part of the FONLL prediction uncertainty band, while non-prompt \dplus{} mesons are in good agreement with the central predictions. A similar behaviour is observed for heavy-flavour decay electrons, where the cross section is on the upper edge of FONLL predictions for \(\pt{} < 5 \gevc{}\) and moves towards the central values at high \pt{}, where beauty-hadron decays are the dominant contribution.

\section{\(\boldsymbol{\ds{}}\)-meson abundance as a function of particle multiplicity}
\begin{figure}[hbt]
	\centerline{
		\includegraphics[height=7.2cm]{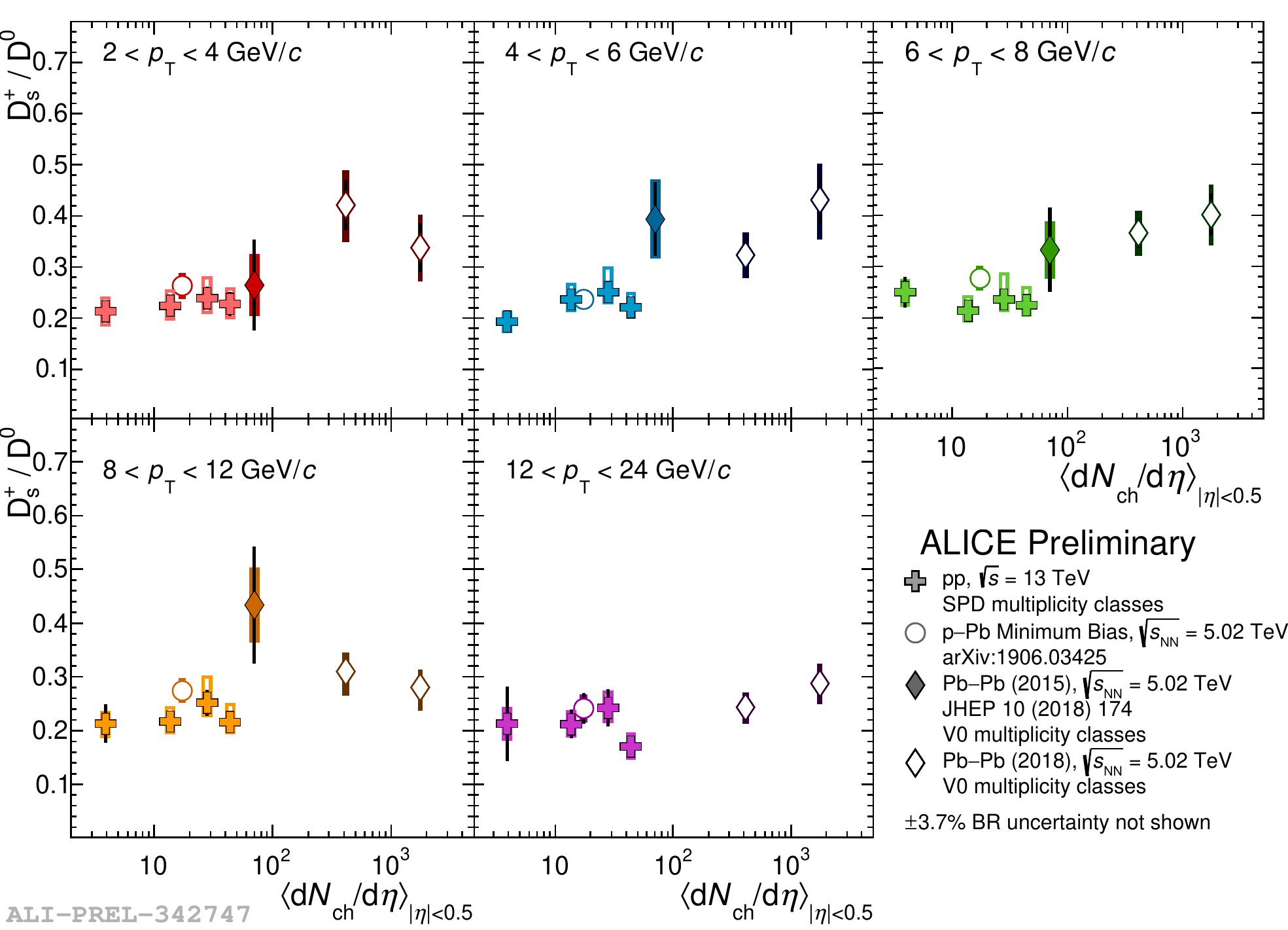}
	}
	\caption{\(\ds{}/\dzero{}\) ratio measured in pp, p--Pb and Pb--Pb collisions as a function of the charged-particle multiplicity and for different \pt{} intervals. \label{dsmult}}
\end{figure}

Figure \ref{dsmult} shows the yield ratio between \ds{} and \dzero{} mesons measured by the ALICE experiment in p--Pb \cite{Acharya:2019mno} and Pb--Pb \cite{Acharya:2018hre} collisions at \cmetnn{}, as a function of the charged-particle multiplicity for different \pt{} intervals, together with new measurements in pp at \cmetthirteen{} and in Pb--Pb collisions.
The \(\ds{}/\dzero{}\) ratio in pp collisions does not show a dependence on the event multiplicity and it is in agreement with the expected value considering the charm-quark fragmentation fractions measured in \(\mathrm{e}^+\mathrm{e}^-\) collisions at LEP \cite{Gladilin:2014tba}. The measurement in minimum bias p--Pb collisions is compatible to what is observed in pp collisions at similar multiplicity values. An increase of the \(\ds{}/\dzero{}\) ratio with respect to pp and p--Pb is observed in Pb--Pb collisions at \(\pt{} < 8 \gevc{}\). This higher production of \ds{} mesons in Pb--Pb is expected if charm quarks hadronise via coalescence with a quark of the QGP medium, where the production of \(\mathrm{s}\bar{\mathrm{s}}\) pairs is enhanced \cite{Kuznetsova:2006bh}.

\section{Open heavy-flavour nuclear modification factor}
\begin{figure}[hbt]
	\centerline{
		\includegraphics[height=5.8cm]{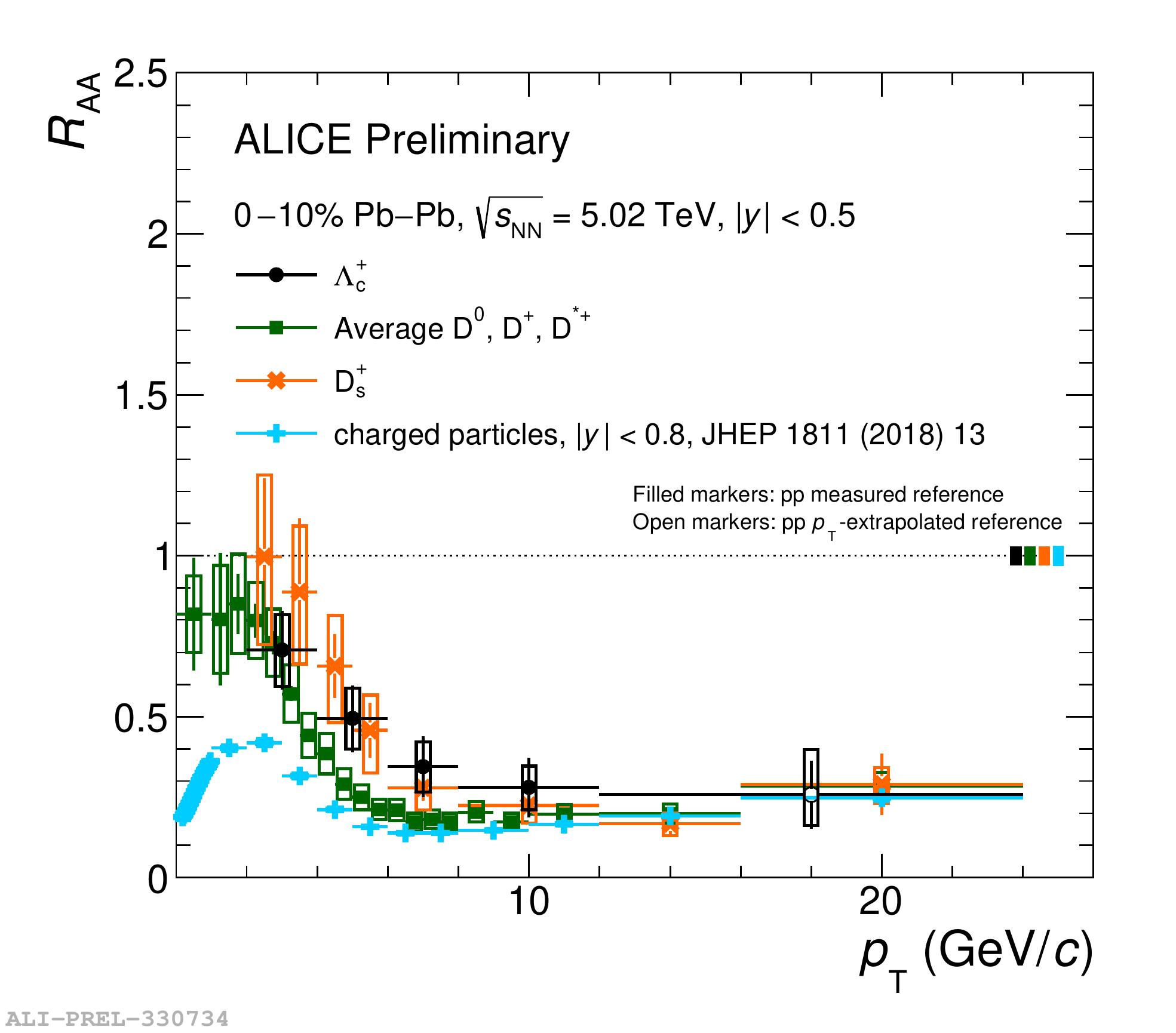}
    \includegraphics[height=5.8cm]{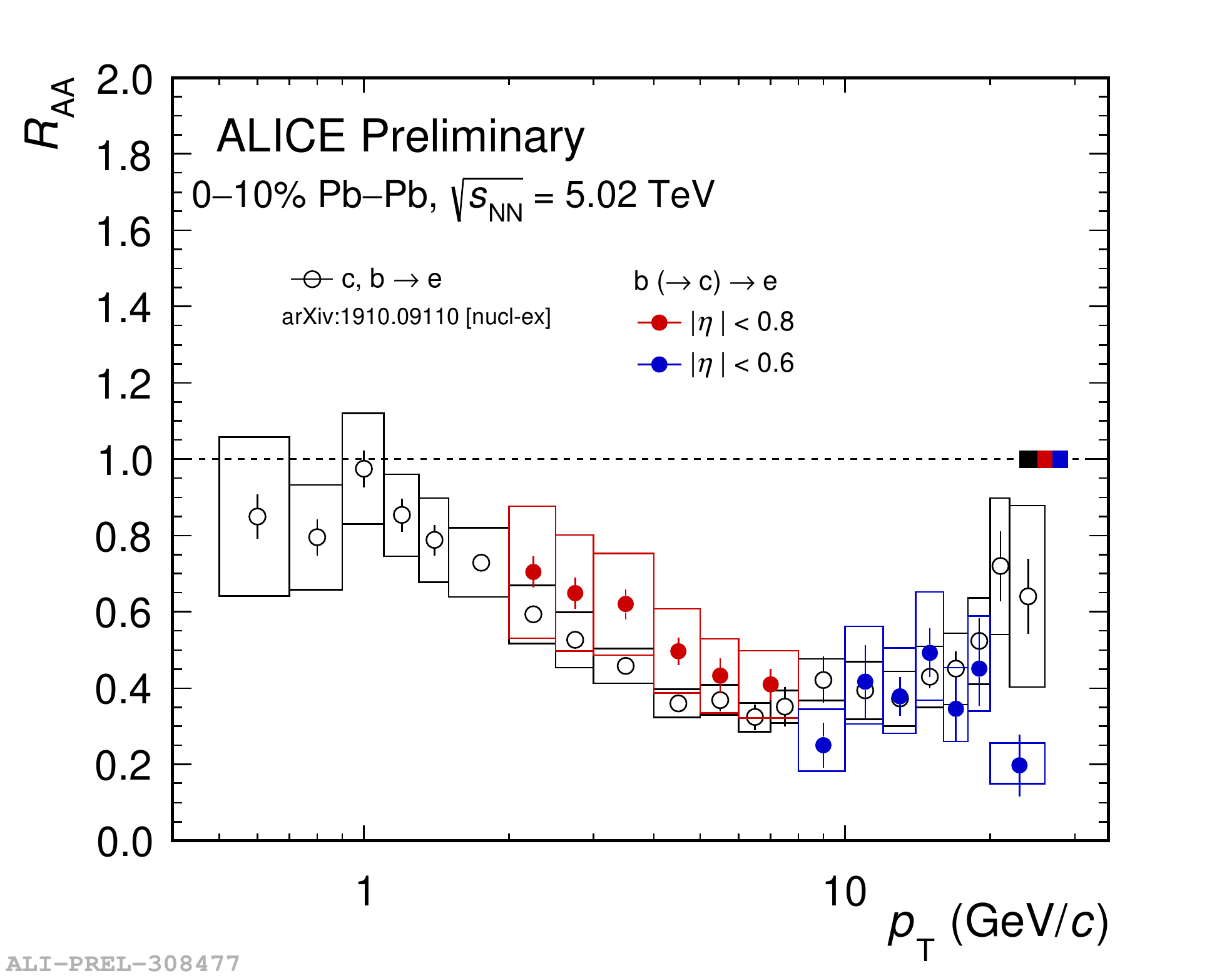}
	}
	\caption{Left: \raa{} comparison between \lc{}, non-strange D mesons, \ds{} and charged particles in the 0--10\% centrality class. Right: \raa{} comparison of \epm{} from heavy-flavour hadron and beauty-hadron decays. \label{raa}}
\end{figure}

The production of prompt charmed hadrons and electrons from heavy-flavour hadron decays \cite{Acharya:2019mom} is measured in Pb--Pb collisions at \cmetnn{} and compared to pp collisions through the nuclear modification factor \(\raa \ (\pt) =  (\mathrm{d}N_{\mathrm{AA}} / \mathrm{d}\pt) / (\langle N^{\mathrm{AA}}_{coll} \rangle \cdot \mathrm{d}N_{\mathrm{pp}} / \mathrm{d}\pt)\); where \(\mathrm{d}N_{\mathrm{AA}} / \mathrm{d}\pt\) and \(\mathrm{d}N_{\mathrm{pp}} / \mathrm{d}\pt\) are the \pt{}-differential yields measured in nucleus--nucleus and pp collisions, respectively, and \(\langle N^{\mathrm{AA}}_{coll} \rangle\) is the average number of binary interactions in a nucleus--nucleus collision. In the left panel of Fig.~\ref{raa}, the measured \raa{} of strange and non-strange D mesons, \lc{} and charged particles \cite{Acharya:2018qsh} in central Pb--Pb collisions are reported. 
A strong suppression of the charmed and light-flavour hadron \raa{} is observed, as expected in the presence of the QGP due to in-medium energy loss. The \raa{} of non-strange D mesons is higher than that of charged particles below \(4 \gevc{}\), while they are compatible at higher \pt{}. This behaviour can be explained by the mass and colour-charge dependence of energy loss. However, also other factors play a role, such as the different initial \pt{} distributions and fragmentation functions of charm and light quarks and the different effects of hadronisation via recombination and radial flow \cite{Acharya:2018hre}. Finally, there is an indication of a smaller suppression of \ds{} mesons and \lc{} baryons than non-strange D mesons at \(\pt{} < 8 \gevc{}\) . In the right panel of Fig.~\ref{raa}, the \raa{} of electrons from semi-leptonic beauty-hadron decays in central Pb--Pb collisions is compared to that of inclusive heavy-flavour decay electrons. The \raa{} of electrons coming from beauty hadrons is above the inclusive one in all the \pt{} intervals below \(10 \gevc{}\), even if compatible within uncertainties, pointing to a possible mass dependence of heavy-quark energy loss, where beauty quarks lose less energy in the medium than charm quarks.

\section{Conclusions}
The ALICE collaboration has measured the production of charmed hadrons and electrons from heavy-flavour hadron decays in different collision systems and at different centre-of-mass energies. In addition, the production of beauty quarks has been investigated with the measurement of non-prompt D mesons and electrons from beauty-hadron decays.

In pp collisions, the measured D-meson and heavy-flavour electron cross sections are compatible with pQCD calculations. In Pb--Pb collisions, the indication of a higher \(\ds{}/\dzero{}\) ratio than in pp, at low \pt{}, is in agreement with the charm-quark hadronisation via quark recombination in the QGP. The D-meson \raa{} is higher than that of charged particles at \(\pt{} < 4 \gevc{}\), suggesting a colour-charge and quark-mass dependence of the energy loss. Furthermore, an indication of a smaller energy loss of beauty quarks than charm quarks is observed through the measurement of electrons from beauty-hadron decays.

\bibliographystyle{utphys}
\small
\bibliography{references}

\end{document}